# Integrating Generative AI in BIM Education: Insights from Classroom Implementation

Islem Sahraoui, Kinam Kim, Ph.D.*, Lu Gao, Ph.D., Zia Din, Ph.D., and Ahmed Senouci, Ph.D.

Department of Civil and Environmental Engineering,
University of Houston,
Houston, TX, United States
*Corresponding author

**Acknowledgement**

The authors acknowledge support from the Office of the Provost at the University of Houston for this research (AI-Powered Building Information Modeling as an Interactive Learning Platform).

# Integrating Generative AI in BIM Education: Insights from Classroom Implementation

**Abstract**

This study evaluates the implementation of a Generative AI-powered rule-checking workflow within a graduate-level Building Information Modeling (BIM) course at a U.S. university. Over two semesters, 55 students participated in a classroom-based pilot exploring the use of GenAI for BIM compliance tasks, an area with limited prior research. The instructional design included lectures on prompt engineering and AI-driven rule checking, followed by an assignment where students used a large language model (LLM) to identify code violations in designs using Autodesk Revit. Surveys and interviews were conducted to assess student workload, learning effectiveness, and overall experience, using the NASA-TLX scale and regression analysis. Findings indicate students generally achieved learning objectives but faced challenges such as difficulties debugging AI-generated code and inconsistent tool performance, probably due to their limited prompt engineering experience. These issues increased cognitive and emotional strain, especially among students with minimal programming backgrounds. Despite these challenges, students expressed strong interest in future GenAI applications, particularly with clear instructional support.

## Introduction

### *Digital Transformation in Architecture, Engineering, and Construction (AEC) and Education*

The AEC industry is experiencing a profound and rapid digital transformation, entering what many describe as a "transformative era" (Rane et al., 2024). This shift is primarily driven by the imperative to driven by the need to integrate advanced technologies to enhance efficiency, foster collaboration, and improve project outcomes(Rane et al., 2023c, 2024). A central component of this shift is the widespread adoption of BIM, which is now integral to how buildings are conceptualized, constructed, and managed (Zawada et al., 2024).

This transition is further supported by government incentives and industry-wide demand for more efficient digital workflows (Rocha & Mateus, 2021). For instance, a UK and Ireland survey revealed that BIM usage surged from 13% to 73% in 2020, with another 26% planning future adoption(Rocha & Mateus, 2021). In North America, for instance, 53.6% of professionals report using BIM in at least three-fifths of their projects, and some firms attribute over 80% of their revenue to BIM-related work (Rocha & Mateus, 2021). Such adoption trends signal that BIM has become a global construction standard and a driving force behind the industry's transition into a digital era comparable to Industry 4.0 (Zawada et al., 2024).

This accelerating digitalization of AEC practice places growing pressure on educational institutions to adapt. Curriculum reform is increasingly viewed as a pedagogical necessity, as tomorrow's professionals must be prepared not only to operate these digital tools but also to engage with evolving industry practices (Dang et al., 2023; Song et al., 2022; Wang et al., 2020). In response, many architecture and construction programs have integrated BIM instruction, often using tools such as Autodesk Revit and Navisworks to teach concepts such as design coordination, clash detection, process virtualization, and Virtual Design and Construction (VDC)(Dang et al., 2023; Zawada et al., 2024)

*The Rise of Generative AI (GenAI)*

Building upon advancements of GenAI, particularly LLMs such as ChatGPT, represents the next frontier of innovation in AEC education (Rane, 2023a). Their capacity to produce novel, diverse, and realistic content based on given prompts is central to their transformative potential within the AEC sector (Taiwo et al., 2024). These advanced AI tools offer significant capabilities for interpreting complex building codes and automating rule-checking. Generative models can analyze building codes and regulations to identify relevant requirements, subsequently generating concise, customized reports for architects (Onatayo et al., 2024a). Furthermore, LLMs can generate scripts or code to automate various engineering and construction processes (Onatayo et al., 2024b).

Beyond design and documentation, the promise of GenAI also extends to supporting cognitively demanding tasks such as compliance checking. These tools can automate complex activities, including construction scheduling, hazard recognition, and resource levelling (Samsami, 2024). By analyzing extensive textual data from blueprints and technical documents, AI provides invaluable insights to architects, engineers, and project managers (Onatayo et al., 2024a). While AI significantly enhances efficiency and accuracy, particularly when prompt engineering is done well these tools are meant to augment rather than replace human expertise, supporting complex decision-making through a collaborative, human-in-the-loop model (Samsami, 2024).

*Challenges and Research Gap*

Traditionally, rule-checking applications in BIM remain heavily reliant on manual programming, which poses additional barriers to students lacking programming experience. Manual methods are often time-consuming, error-prone, and difficult to scale, with rule information hard-coded in a way that limits user interaction and adaptability (Aimi Sara Ismail et al., 2017; Chen et al., 2024). Even semi-automated systems often require substantial manual input, making them inaccessible to those without technical expertise (Chen et al., 2024).

In addition, despite growing enthusiasm for AI in education, there is still a significant lack of research on how GenAI is perceived and used by students in real classroom settings. Much of the literature focuses on theoretical opportunities or tool development, leaving a gap in understanding actual instructional outcomes and student experiences (Chan & Hu, 2023). This is particularly important in applied fields such as construction management, where tool usability, instructional scaffolding, and student adaptability play a major role in successful implementation.

Moreover, current AI education strategies face systemic limitations. Many AI-focused courses require coding skills, which can deter participation and affect long-term knowledge retention if not followed up with continuous engagement (Cheng et al., 2025). Instructors themselves often report uncertainty around AI integration and cite a lack of expertise as a key barrier, even as they acknowledge the potential benefits of AI tools in teaching and learning (Cheng et al., 2025). These concerns are echoed by scholars calling for more effective pedagogical models, clearer guidance on integrating AI into curricula, and a stronger emphasis on validating the accuracy and reliability of AI-generated outputs in educational environments (Peláez-Sánchez et al., 2024).

*Research Questions and Objectives*

To address these educational and technological shifts, this study investigates the implementation of GenAI within graduate-level BIM-related courses. The objective is to evaluate how such tools can be pedagogically integrated to support learning, and what implications arise from their use in a structured academic environment. Specifically, the study seeks to answer three core research questions: (1) How do students experience and perceive the use of GenAI tools such as LLMS for an automated rule-checking framework in BIM? (2) What specific technical and cognitive challenges do students face when interacting with these tools, particularly in tasks involving prompt engineering, debugging, or interpreting AI-generated outputs? (3) From an instructional standpoint, how effective is the integration of GenAI, and what design strategies can enhance its educational value?

To explore these questions, a mixed-methods approach was employed across two semesters. The methodology combines quantitative data such as NASA-TLX workload metrics and assignment performance scores with qualitative insights gathered through student perception surveys and post-assignment interviews. This multi-pronged approach enables a more holistic understanding of the student learning experience, capturing not only measurable outcomes but also the emotional and cognitive dimensions of engaging with AI. By triangulating these data sources, the study aims to identify practical lessons learned, instructional gaps, and support mechanisms needed for successful integration. Ultimately, the goal is to inform future curriculum design by offering empirical guidance on how to responsibly and effectively incorporate GenAI into BIM education and similar digitally intensive domains.

The remainder of this paper is structured as follows. The literature review section delves into the adoption and application of GenAI within the construction industry and educational environments. This is followed by the methodology, which outlines the research design, data collection methods, and analytical approaches utilized. The results section presents and analyzes the study's findings, incorporating both quantitative data and qualitative insights derived from student experiences. The paper subsequently discusses the implications for practice and pedagogy and concludes with recommendations for future research and curriculum development in the realm of AI-driven BIM education.

This paper provides new empirical insights into using GenAI in construction education, addressing the need for effective instructional models and support. It highlights student experiences, outcomes, and key lessons, offering practical recommendations for integrating AI into BIM and construction management curricula.

**Literature Review**

*GenAI in the Construction Industry*

The AEC industry is witnessing the nascent but growing integration of Generative Artificial Intelligence (GenAI), characterized by its evolving adoption trends and the emergence of diverse practical applications across the project lifecycle. GenAI has emerged as a powerful tool capable of learning from and creatively utilizing existing data to generate new design ideas and insights within these sectors, with its development and application, particularly in building design, showing significant progress over the past three years (Ghimire et al., 2024). LLMs, including prominent examples such as OpenAI's GPT, Google's PaLM, and Meta's Llama, are at the forefront of this transformation, demonstrating substantial potential and global interest in the construction industry (Ghimire et al., 2023, 2024). These models, built upon transformer architectures, are specifically designed for natural language processing (NLP) tasks, enabling them to generate human-like text and integrate into various AI-driven content creation processes (Wan et al., 2025). This transformation extends beyond formal construction sectors; recent work highlights how AI can also enhance informal construction practices by supporting decision-making, improving labor conditions, and enabling more sustainable development strategies (Cédric Cabral et al., 2022). Current adoption rates indicate that many AEC professionals are in the early stages of integrating these tools, with a LinkedIn poll revealing that 40% have never used GenAI, and most companies lack formal guidelines on its usage (Ghimire et al., 2023, 2024). Despite this nascent stage, LLMs are increasingly being implemented to enhance productivity, accuracy, and automation across diverse construction tasks (Taiwo et al., 2024). AI should be treated as a collaborative tool that empowers professionals and learners to shift from routine tasks to strategic, higher-order thinking

(Sharma & Gupta, 2025).

These implementations include GenAI's facilitation of intricate design concept generation and enhanced collaboration among architects, engineers, and stakeholders, accelerating the creation of detailed BIM models (Rane, 2023b), notably through BIM-GPT, a prompt-based virtual assistant framework that integrates LLMs like ChatGPT with BIM systems. It enables users to interact with BIM data through natural language prompts, allowing them to retrieve information, generate summaries, and answer queries about building components. The framework includes a prompt library, a prompt manager, a natural language interface, and a visualization module to support intuitive and context-aware interaction with complex 3D BIM models (Zheng & Fischer, 2023). Furthermore, GenAI learns from existing design data and empirical rules to intelligently generate building structures to improve efficiency and accuracy for safer and more sustainable designs (Ghimire et al., 2023). LLMs also can automate repetitive tasks in construction planning and management, such as schedule generation, demonstrating the ability to create logical sequences and meet project requirements, and aiding in risk assessment by identifying potential risks and supporting contingency planning through analysis of project data and historical trends (Onatayo et al., 2024a; Rane, 2023b; Taiwo et al., 2024). AI-BIM frameworks have also proven valuable in infrastructure monitoring, supporting real-time visualization, planning optimization, and intelligent decision-making throughout project life cycles (Yang & Xia, 2023).

In automation and robotics, LLMs like ChatGPT are leveraged for automated sequence planning in robot-based construction assembly, as demonstrated by RoboGPT, a novel robotic system developed to automate sequence planning for complex construction assembly tasks. Its core logic leverages ChatGPT's advanced reasoning capabilities to understand task requirements and generate step-by-step sequential commands for robot execution, such as assembling components or installing systems. The system integrates a robot control system, a scene semantic system for object detection, and a command decoder to translate ChatGPT's natural language responses into machine-understandable commands. While RoboGPT significantly automates complex construction sequence planning, it currently faces limitations in precise spatial understanding from text input (You et al., 2023). LLMs can also streamline operations by generating automated workflows and strategies for controlling robotic systems (Rane, 2023b). GenAI enables the automated synthesis of construction documents and facilitates question-answering systems, unlocking valuable information from dense, unstructured contract documents (Ghimire et al., 2023, 2024; Samsami, 2024), while also aiding in preparing proposals, bids, and daily progress reports (Onatayo et al., 2024a; Taiwo et al., 2024).

Despite the transformative potential of GenAI and LLMs, their widespread adoption and effective integration in the construction industry are impeded by significant multi-faceted challenges (Taiwo et al., 2024). A key challenge lies in the domain-specific knowledge required for construction's diverse disciplines (structural, mechanical, electrical, plumbing, project management), as GenAI models often rely on statistical patterns and lack the explicit encoding of intricate technical knowledge necessary to extract meaningful relationships from industry data (Onatayo et al., 2024a). Furthermore, inherent model limitations pose considerable hurdles, including accuracy issues and "hallucinations" where LLMs produce convincing but false or inaccurate outputs due to limited knowledge or noisy training data, posing risks for critical construction decisions (Ghimire et al., 2024; Onatayo et al., 2024a; Samsami, 2024). Models often exhibit poor generalizability, struggling to perform well beyond their specific training datasets, which makes their application challenging in diverse real-world construction environments (Taiwo et al., 2024). Moreover, the dynamic nature of construction materials, methods, and regulations necessitates continuous and costly retraining of AI models to prevent them from becoming outdated and providing unreliable guidance (Samsami, 2024). The substantial computational demands for training and operating GenAI models also present significant cost barriers, particularly for smaller construction companies (Wan et al., 2025). Beyond model-specific issues, broader adoption barriers include resistance to change and skepticism about AI's reliability prevalent within the industry (Samsami, 2024), coupled with a noted lack of specialized AI skills and expertise among most construction companies, necessitating a skilled workforce proficient in AI technologies (Rane, 2023a). Finally, current studies reveal an over-reliance on specific LLMs such as ChatGPT, with limited exploration of open-source alternatives and a lack of quantitative comparisons across different models and against traditional human efforts, indicating a need for more robust comparative research (Wan et al., 2025).

*GenAI in Higher Education*

The integration and adoption of GenAI and LLMs are profoundly reshaping Higher Education (HE) and, more specifically, the AEC sector. The release of open GenAI tools like ChatGPT in 2022 significantly accelerated scholars' interest in their educational impact, raising societal awareness of AI's existence and potential (Alhafni et al., 2024; Kaplan-Rakowski et al., 2023). Teachers generally view GenAI positively, with over 80% of surveyed teachers considering ChatGPT a "valuable instructional tool for students" (Kaplan-Rakowski et al., 2023), and 54% of teachers indicated they are likely or very likely to integrate AI into their teaching (Kaplan-Rakowski et al., 2023). This readiness suggests that teachers are willing to learn and adapt to integrate these technologies, partially because GenAI tools like ChatGPT are relatively easy to use with fewer external barriers than previous technologies (Kaplan-Rakowski et al., 2023). Similarly, university students generally hold a positive attitude towards GenAI and express willingness to integrate it into their learning and future careers (Chan & Hu, 2023). Within the AEC sector, educators recognize AI literacy as a crucial computing-related competency for students to develop in the next 5–10 years (Cheng et al., 2025). Despite this potential, the adoption and integration of AI into AEC education are still considered nascent and largely underexplored in curricula(Cheng et al., 2025; Komatina et al., 2024). For instance, 65% of construction faculty reported never having integrated AI content into their courses, with this figure rising to 77% for U.S. respondents compared to 48% for international respondents (Cheng et al., 2025). In addition, a considerable portion of staff felt inadequately prepared and desired further support (Lee et al., 2024).

  Beyond its current status, GenAI presents numerous opportunities to significantly enhance the educational experience through personalized support, advanced assessment tools, and improved creative and technical problem-solving capabilities. A primary benefit lies in its capacity to generate original output in response to user prompts, fostering a generally positive student attitude towards its application in teaching and learning (Chan & Hu, 2023). A survey conducted with 399 undergraduate and postgraduate students from various disciplines in Hong Kong revealed a generally positive attitude among students towards GenAI in teaching and learning, with their willingness to use GenAI correlating positively with their knowledge and frequency of use. Students recognized GenAI's potential for personalized and immediate learning support, personalized feedback, and time-saving benefits. perceived usefulness and ease of use drive intention to use, while trust and ethical concerns moderate sustained engagement(Chan & Hu, 2023). In experimental settings, a study showed that students who used ChatGPT to solve creative tasks felt more confident, performed better, and expended less mental effort. However, students also struggled with accurate self-assessment, highlighting the need for metacognitive training to guide AI-supported learning. While engagement was high, it was driven partly by the perceived ease of using AI, potentially masking over-reliance risks (Urban et al., 2024).

Furthermore, GenAI tools are highly effective in providing personalized learning support, offering tailored tutoring and immediate feedback that adapts to individual student needs and progress. A study proposes a framework for guiding the use of GenAI in education, discusses its potential benefits, limitations, and future directions. It suggests that ChatGPT can create virtual tutors, answer student questions, and provide personalized learning experiences by adjusting the learning level and pace. The framework supports meaningful engagement by guiding GenAI use around clearly defined learning outcomes and ongoing evaluation of educational effectiveness (Su & Yang, 2023). Another study has demonstrated that GenAI also serves as a valuable assessment and teaching aid, automating essay scoring and grading to deliver consistent and immediate feedback to students, which streamlines the assessment process (Yu & Guo, 2023). Educators can leverage GenAI to assist in developing syllabi, quizzes, exams, and lesson plans, consequently reducing workload and enhancing overall teaching practices (Baidoo-Anu & Owusu Ansah, 2023; Lo et al., 2024; Ogunleye et al., 2024). Moreover, GenAI tools, including LLMs as ChatGPT, are specifically recognized for their capability to create new content across various formats, such as high-quality articles, and the ability to both write and debug original computer code, which makes them versatile assets for both creative and technical tasks (Baidoo-Anu & Owusu Ansah, 2023; Yu & Guo, 2023; Yusuf et al., 2024). These capabilities offer significant potential for students and educators in various subject areas, particularly those involving content creation or programming.

Although AI adoption is expanding across the construction industry, its integration into construction education remains limited. Several sources describe this adoption as still in its infancy; it is also noted that there is an absence of discussion about the integration of AI topics into construction education despite its increasing adoption in the construction field. The integration of GenAI in education is accompanied by several notable challenges and concerns that necessitate careful consideration. A primary concern is the phenomenon of accuracy issues, often termed "hallucinations," where AI models produce plausible but factually incorrect or misaligned information (Baidoo-Anu & Owusu Ansah, 2023; Lee et al., 2024; Ogunleye et al., 2024). Critically, AI tools lack the inherent ability to assess content validity, necessitating continuous human oversight to ensure informational integrity (Chan & Hu, 2023). Another significant challenge pertains to the risk of over-reliance and its potential impact on skill development, as excessive dependence on GenAI tools could impede students' genuine efforts to cultivate essential academic and cognitive skills, including writing competence, critical thinking, problem-solving, and overall intellectual growth (Chan & Hu, 2023; Lo et al., 2024; Yusuf et al., 2024). Furthermore, profound bias, privacy, and ethical issues are inherent to GenAI use, with concerns existing regarding the perpetuation of biases from training data, the privacy of user information, and broader ethical implications surrounding the responsible deployment of these tools in educational settings (Peláez-Sánchez et al., 2024; Su & Yang, 2023; Yu & Guo, 2023; Yusuf et al., 2024). Hence, the emergence of GenAI exacerbates the digital divide and equity concerns, as not all students possess equitable access to advanced GenAI technologies, raising questions about fairness and potential disparities in access to high-quality educational programs, particularly concerning paid LLMs like ChatGPT-4 (Peláez-Sánchez et al., 2024).

*BIM in AEC Education*

The integration of BIM into AEC education programs aims to bridge the gap between academic instruction focused on theoretical principles and the industry's demand for practical, application-based skills (Bozoglu, 2016). As such, many universities have progressively incorporated BIM into their courses and curricula at both undergraduate and graduate levels (Dang et al., 2023). However, a recognized "skill gap" persists between academic construction programs and the actual needs of the construction industry, evidenced by the lack of a unified educational program to foster BIM professionals in the United States (Song et al., 2022). This disparity leads companies to often prefer to hire experienced BIM professionals over recent college graduates for BIM staffing (Song et al., 2022).

  The teaching of BIM extensively utilizes various pedagogical techniques aimed at fostering comprehensive understanding and practical skills. Project-based learning (PBL) is a predominant method, immersing students in realistic projects where they apply BIM concepts, which enhances motivation and learning effectiveness (Besné et al., 2021). This often goes hand-in-hand with experiential learning, emphasizing "learning by doing" to cultivate changes in judgment, feeling, and skills (Bozoglu, 2016). Collaborative learning and group work are central, mirroring industry practices by promoting effective communication and teamwork through interdisciplinary design and coordination tasks, sometimes even necessitating classroom layout changes to simulate integrated design environments (Meterelliyöz & Özener, 2022). While traditional teacher-led instruction through lectures and lab tutorials conveys foundational concepts, guided self-learning is encouraged via web-based video tutorials and online resources, allowing lab sessions to focus on problem-solving. Videos are often preferred over text-based handouts for their ability to support PBL and increase student engagement (Nguyen et al., 2025; Zulfikar A. Adamu & Tony Thorpe, 2016). Additionally, just-in-time and asynchronous instruction methods provide tailored support for common problems (Meterelliyöz & Özener, 2022).

The range of technologies and tools used in BIM education is diverse, designed to provide students with hands-on experience and prepare them for industry demands. Core BIM software includes widely utilized platforms like Autodesk Revit for 3D modeling in architectural, structural, and MEP disciplines, as well as for 4D (scheduling) and 5D (costing) applications (Nguyen et al., 2025). Other essential tools include Autodesk Navisworks for clash detection, model auditing, coordination, and foundational computer-aided design tools like AutoCAD (Dang et al., 2023).

Beyond fundamental BIM software skills, AEC programs are increasingly incorporating cutting-edge technologies and advanced concepts. For example, a study proposes a conceptual framework for integrating BIM with LLMs such as ChatGPT to enhance AEC workflows. The framework envisions a bidirectional interaction where BIM data is used to enrich the prompt context, enabling the LLM to generate more accurate, domain-specific responses. In turn, ChatGPT interprets natural language queries and provides insights or textual descriptions based on BIM components, such as identifying room functions or suggesting material specifications. This integration leverages structured BIM data as input and utilizes natural language output to improve user interaction, collaboration, and design decision-making. The study emphasizes that enriching LLMs with BIM content allows for a more intelligent assistant that understands AEC-specific terminology and can support early-stage design ideation and communication among stakeholders (Rane et al., 2023c).

Similarly, Augmented Reality (AR) and Virtual Reality (VR) tools are gaining popularity, being linked to BIM models to provide more immersive and interactive learning experiences (Lozano-Galant et al., 2024). These tools are instrumental in enhancing the visualization of complex engineering concepts and improving student understanding and engagement in civil engineering courses, with AR applied in design studios, interior design, and for understanding specific building components, and VR-embedded BIM systems utilized for quantity surveying education (Lozano-Galant et al., 2024).

Despite this advancement in BIM education and the integration of new technologies, significant challenges persist, primarily related to the complexity of tools, a persistent skills gap, and issues with technology standardization and interoperability. The inherent complexity of BIM tools and their industry-centric interfaces can pose difficulties for students, potentially leading to slower learning curves (Meterelliyöz & Özener, 2022). There is also a recognized disconnect between fundamental computer science concepts and current educational approaches, underscoring a need for improved learning strategies (Meterelliyöz & Özener, 2022). Many current BIM courses still primarily focus on theoretical concepts or basic modeling skills (Song et al., 2022), contributing to a significant gap between academic training and industry expectations, particularly concerning graduates' readiness for immediate project deployment (Song et al., 2022). Furthermore, the lack of standardized protocols for integrating new technologies like GenAI with BIM presents a challenge to widespread industry adoption (Rane et al., 2023c). The continuous evolution of both BIM and GenAI necessitates proactive system maintenance and updates to ensure relevance and effectiveness. Interoperability issues between different BIM software and the complex interpretation of multi-faceted data also remain ongoing hurdles (Rane et al., 2023c).

*Automated Rule-Checking in BIM*

Ensuring compliance with complex industry standards and regulations during construction project design and implementation phases presents a significant challenge within the BIM domain, historically addressed by manual, often inefficient, and error-prone methods. Traditional manual compliance checking methods often prove inefficient, time-consuming, and susceptible to errors, failing to meet modern engineering demands, particularly as projects escalate in size and regulatory complexity (Aimi Sara Ismail et al., 2017; Chen et al., 2024; Peng & Liu, 2023). To overcome these limitations, researchers have studied Automated Rule Checking (ARC) systems (Aimi Sara Ismail et al., 2017; Chen et al., 2024). The general structure of an automated code compliance checking system typically involves four main stages: (1) rule interpretation (translating natural language rules into a computer-processable format), (2) building model preparation (gathering necessary information from the model), (3) rule execution (applying computer-processable rules), and (4) reporting results (Aimi Sara Ismail et al., 2017; Chen et al., 2024). The rule interpretation stage is particularly vital and complex, and most early ARC systems relied on manual rule interpretation, making the maintenance and modification of hard-coded rules inefficient (Aimi Sara Ismail et al., 2017; Chen et al., 2024).

To address the complexities of rule interpretation within AEC systems, various techniques have been developed and applied, ranging from commercial software and plug-ins to more sophisticated object-based and ontological approaches. Existing software and plug-in applications, such as Solibri Model Checker (SMC), are prominent in the BIM software landscape, assisting in visualizing design issues and performing automated analysis against built-in geometry-oriented rules (Aimi Sara Ismail et al., 2017; Hjelseth, 2015). While SMC provides user-friendly visualization and has been applied in significant projects, its reliance on hard-coded functions limits the ease of adding new rules (Aimi Sara Ismail et al., 2017; Chen et al., 2024). The object-based approach organizes knowledge by representing object types, defining attributes, procedures, rules, and machine learning elements, exemplified by Singapore's CORENET e-PlanCheck and Australia's DesignCheck system, which categorize rules based on descriptions, performance requirements, objects, properties, and relationships (Aimi Sara Ismail et al., 2017). An ontological approach leverages semantic web technologies to overcome limitations of Industry Foundation Classes (IFC) in automated code compliance checking by formally defining concepts and relationships within a domain (Aimi Sara Ismail et al., 2017; Peng & Liu, 2023). Despite progress with these traditional methods, including rule-based and machine learning approaches, challenges persist, such as the need for extensive manual feature engineering, large annotated datasets, and significant computational resources (Chen et al., 2024).

      The advent of AI technologies, including NLP, deep learning (DL), and LLMs, has significantly advanced automated code compliance checking by directly addressing many of the limitations inherent in earlier approaches. NLP techniques are widely used for processing and understanding human language-based text to automate rule extraction from regulatory documents, typically involving information extraction (semantic information from text) and information transformation (converting extracted information into logical clauses) (Chen et al., 2024). However, traditional NLP methods often require extensive manual feature engineering and large amounts of manually annotated training data, demanding substantial technical investment for high accuracy (Chen et al., 2024). Deep learning models circumvent the need for manually defined features by automatically extracting and learning features through multi-layer neural networks, simplifying traditional multi-stage processing through an end-to-end learning

approach (Chen et al., 2024). Recurrent Neural Networks (RNNs) have been used to extract hierarchical information from building regulations, and BERT (Bidirectional Encoder Representations from Transformers) demonstrates strong performance in complex text classification due to its bidirectional training structure and deep contextual understanding (Chen et al., 2024).

Nevertheless, DL models generally require larger datasets and substantial computational resources for training, a challenge in the AEC field's data scarcity, although transfer learning can mitigate this (Chen et al., 2024). Their internal complexity also poses interpretability issues (Chen et al., 2024). LLMS, such as OpenAI's GPT series, Meta's Llama models, and Claude, offer robust language understanding with minimal labeled data due to extensive pre-training on diverse corpora and few-shot learning capabilities (Chen et al., 2024), significantly reducing the need for large, annotated datasets required by previous methods. LLMs can adapt to evolving regulations, accurately extract structured information from regulatory texts, and exhibit "emergent capability" to handle unseen data, reducing development and maintenance costs and expanding application ranges (Chen et al., 2024). However, their application in the AEC field is still limited by the need for fine-tuning for specific tasks and challenges in handling complex texts with nested clauses and conditional statements (Chen et al., 2024).

*Challenges and Research Gaps*

Despite the transformative potential of GenAI, its integration into BIM education faces significant challenges that this study aims to investigate and address, particularly concerning the accessibility of existing methods and a notable research gap in classroom implementation. Most existing studies on GenAI in education tend to focus on AI development, testing of model capabilities, and opportunities for applications in education, with limited empirical research specifically dedicated to classroom implementation and its instructional outcomes (Ogunleye et al., 2024). Traditional rule-checking instruction often relies on manual coding, which is inherently time-consuming and largely inaccessible for students without prior programming experience (Chen et al., 2024). Furthermore, even more advanced methods incorporating AI typically demand substantial expertise for model training and maintenance, posing a high barrier to entry for educators and learners alike(Lo et al., 2024; Su & Yang, 2023).

Recent literature further amplifies these concerns, introducing new dimensions to the challenges of integrating GenAI in BIM education and practice. Ethical issues, such as ensuring data privacy and mitigating the risk of student over-reliance on AI outputs, introduce complexities that could potentially undermine critical thinking and independent problem-solving (Yu & Guo, 2023). This paper seeks to contribute empirical insights into these pedagogical and technical challenges through the practical implementation and evaluation of an AI-powered rule-checking workflow in a BIM educational setting, directly informing future instructional design and support strategies.

**Methodology**

This study evaluates the educational impact of integrating AI-powered rule-checking into graduate-level BIM-related courses, with a focus on student adaptation, instructional usability, and integration into existing workflows. The instructional design consisted of three main phases: two lectures introducing manual and AI-based rule-checking techniques, an assignment where students applied GenAI tools for rule verification, and a structured survey to assess student experience and learning outcomes. The overall structure of this methodology is outlined in Figure 1.

The first phase of the workflow focused on BIM model rule-checking, where both manual and AI-driven compliance checking approaches were introduced to students through two lectures, about three hours each. The manual method required students to perform rule lookups from code documents such as the International Building Code (IBC) or National Fire Protection Association (NFPA), review BIM model elements, and manually compare them to the relevant regulatory requirements. This process emphasized the traditional challenges of code compliance verification, including the labor-intensive nature of interpretation, time-consuming processes, and the potential for human error. In contrast, the AI-based method introduced students to LLMs as tools to automate rule-checking tasks. The instruction included an explanation of the capabilities, potential and limitations of LLMs, a practical tutorial on prompt engineering principles, and a demonstration of an AI-driven system, which is a custom workflow that enables rule extraction, code translation, and Python script generation. This comparative instructional design allowed students to understand both foundational and emerging techniques for compliance checking within the BIM environment.

The second phase involved a structured assignment in which students applied the AI-based method to automate compliance verification tasks using GenAI tools. students were assigned a set of five building code rules derived from the IBC or NFPA. They were required to formulate detailed prompts to query an LLM of their choice, with the objective of generating functional Python scripts capable of checking rule compliance within a BIM software, i.e., Revit. After generating the scripts, students executed them in the Revit Python Shell, where the generated scripts can be executed. Errors encountered during the scripting process were used to reinforce prompt refinement and model debugging skills. The final output was a compliance report that documented the violations identified by the AI-generated code and evaluated the reliability and accuracy of the results. This assignment phase not only developed students' technical competencies in prompt engineering and AI-powered programming but also provided insight into the practical limitations of AI-generated outputs in professional workflows.

The third and final phase comprised a comprehensive evaluation of the instructional implementation of the AI-driven method using a combination of quantitative and qualitative measures. A structured survey was administered to all participants, including standardized workload metrics from the NASA Task Load Index (TLX) to assess mental, physical, and temporal demands, perceived performance, effort, and frustration. In addition to the TLX items, the survey incorporated perception-based questions addressing usability, effectiveness, time efficiency, and openness toward AI tools in engineering tasks. To capture deeper insights, open-ended questions and follow-up interviews were conducted, enabling students to reflect on their experience with the AI-based method and provide constructive feedback. The collected data underwent statistical analysis and thematic coding, facilitating a rigorous interpretation of students' adaptation to the technology and the instructional value of the integration. This phase focused on measuring the educational effectiveness of the lecture, assignment structure, and instructional support, providing insights to guide future pedagogical strategies for integrating emerging technologies into BIM education

*Instructional Framework*

*Lecture 1: Manual Rule-Checking and BIM Model Quality*

The objective of the first lecture was to introduce students to the fundamentals of quality control and rule-based compliance verification in BIM environments. The lecture began by presenting the manual method of rule-checking, where students assess model elements against regulatory codes through direct inspection. While this approach reinforces understanding of compliance principles, it is time-consuming, prone to human error, and difficult to scale for large models. These limitations naturally led to a discussion on the growing need for automation in BIM quality assurance. As project complexity increases and data flows become more integrated, relying solely on manual checks becomes impractical. The lecture highlighted the role of automated rule-checking systems in addressing these challenges by offering faster, repeatable, and more consistent verification processes. Current tools such as Solibri, BIMAssure, and SmartReview were briefly introduced, with attention given to their strengths and constraints, particularly the need for structured model data and predefined rule sets.

A key portion of the lecture presented the formal process of rule testing in BIM environments, broken down into five distinct steps. The first step, Rule Identification, involves selecting specific regulatory provisions to be checked, typically from sources like the IBC or NFPA. The second step, Model View Definition, consists of preparing appropriate views of the model that isolate the elements subject to rule enforcement. Next, in the Element Selection, as shown in Figure 2, users identify the model components to which the rules apply (e.g., windows, doors, or rooms). The fourth step, Rule Application, is where each rule is manually verified by comparing model data against the specified code requirements. Finally, in Failure Identification, any elements that violate the rule are flagged, documented, and prepared for revision.

To reinforce these concepts, the lecture included a demonstration of manual rule-checking applied to window egress requirements, referencing NFPA 101-2024, Section 24.2.2.3.3. According to this standard, windows used for emergency egress in residential occupancies must provide a clear opening of at least 24 inches in height and 20 inches in width, with the bottom of the opening located no more than 44 inches above the floor. During the demonstration, students manually inspected selected windows in a Revit model, checking their dimensions and sill height against the rule. When discrepancies were found, such as windows with insufficient opening dimensions or improper sill height, these were recorded, and screenshots were taken to document the non-compliance. This hands-on example helped students experience the practical demands and attention to detail required in manual compliance checking and provided a foundation for comparing this process to the AI-assisted method introduced later in the course.

*Lecture 2: AI-Based Rule-Checking Systems*

Prompt Engineering – Building Effective AI Interactions

The first part of the second lecture introduced students to the fundamentals of prompt engineering as a critical skill for leveraging GenAI in practical settings. This session emphasized that the effectiveness of AI-generated outputs, particularly for rule-checking in BIM, depends heavily on how well users formulate their prompts. The instructor explained five core principles for crafting effective prompts: specifying the task, including context, providing references, evaluating outputs, and iteratively refining the prompt. Each principle was illustrated with examples relevant to the course, focusing on building code interpretation and automation.

    The session highlighted that effective prompts should define the AI's role (e.g., "act as a code compliance expert"), clarify the format (e.g., Python script), and provide sufficient technical context and constraints to guide the response. Students were shown how poorly defined prompts lead to vague or incorrect answers, while well-structured prompts increase the reliability and usability of the AI-generated output. Advanced strategies such as prompt chaining, tree-of-thought prompting, and meta-prompting were introduced to help students manage complex tasks by breaking them into smaller,

more interpretable steps. These techniques provided scaffolding for building robust AI workflows that go beyond single queries.

The session also included interactive in-class examples and discussions where students tested variations of prompts, analyzed the quality of LLM outputs, and identified hallucinations, inconsistencies, or irrelevant results. To help students structure their prompts effectively, the instructor shared a complete example used in the assignment, illustrating how to instruct the AI step-by-step. This example included the initial prompt and follow-up instructions to resolve errors and refine results. The breakdown is presented in Table 1, and it was used as a guide throughout the AI-driven rule-checking assignment.

Demonstration of AI-Based Rule-Checking

The second part of Lecture 2 aimed to provide students with an applied understanding of how GenAI can be used to automate code compliance checking within BIM. The primary objective of this session was to demonstrate the full lifecycle of AI-assisted rule verification from interpreting building codes to generating executable scripts in Revit. The lecture first reviewed the foundational concepts of LLMs, including how these models are trained, how they understand language using transformers and attention mechanisms, and their capabilities in text generation, contextual understanding, and code synthesis. It also outlined key limitations specific to the construction domain, such as the LLMs' lack of direct access to BIM data, their dependence on Revit Python Shell, and the potential for producing errors or hallucinations in code output.

To illustrate the end-to-end process of AI-driven compliance verification, the lecture introduced the workflow shown in Figure 3. This framework begins with input data comprising both the BIM model and regulatory text (e.g., IBC or NFPA rules). The AI-based interpretation module leverages LLMs to perform natural language processing, contextual understanding, and semantic analysis of the rules. From this interpretation, Python scripts are dynamically generated to automate rule-checking tasks. When errors occur, LLMs are prompted with error messages and contextual clarifications to iteratively revise the script. The final outputs include compliance reports and improvement suggestions. This figure provided students with a structured mental model for how GenAI tools integrate with traditional BIM workflows to support compliance verification.

  In the hands-on component of the lecture, students observed as the instructor applied an AI-assisted approach to check several code compliance rules using a Revit model. The demonstration focused on regulatory requirements from IBC 2018 and NFPA 101-2024, including stair riser and tread dimensions, window egress requirements, door clear width, emergency escape openings, and foundation depth. To verify these rules, the instructor used a predefined, well-structured prompt to query an LLM, which then generated Python scripts tailored for compliance checking in Revit. When initial scripts failed due to syntax issues or incomplete logic, the instructor reused the LLM by feeding back the error messages to iteratively refine the code. After successfully executing the scripts, the instructor displayed the resulting compliance report, which identified non-compliant elements within the BIM model. This process allowed students to observe the full AI-based compliance workflow and understand both its potential and its practical limitations in real-world construction applications. To evaluate the implementation of the automated rule compliance checking frameworkin educational settings, students were assigned an individual task involving the use of LLMs to automate the detection of code violations in a Revit model. The assignment was designed to simulate a practical workflow where students acted as both prompt engineers and compliance analysts using AI tools integrated with Autodesk Revit.

Students began by crafting structured prompts to guide a GenAI model (e.g., ChatGPT or Bard) in producing Python scripts. These scripts were designed to test specific building code requirements within the BIM model. The prompting process followed a defined framework, encouraging students to specify the AI's role, provide relevant context, include rule references, evaluate script output, and iteratively refine prompts based on code performance and feedback. Once the script was generated, it was executed in Autodesk Revit 2024 via the Revit Python Shell. The script automatically scanned model components and flagged those not complying with the rule parameters. Students identified the corresponding elements in the model using the element IDs returned by the script, visually inspected them, and documented their findings. Each student completed a results table that included the rule tested, the final prompt version, screenshots of the flagged elements, and a concise explanation of each violation and its potential implications.

The rules tested in the assignment were drawn from IBC 2018 and NFPA 101-2024, chosen for their relevance to measurable and automatable aspects of BIM modeling. These included parapet height (minimum 30 inches above roof intersection, Section 705.11.1), corridor width (at least 44 inches for main corridors, Section 1020.2), and ceiling height (minimum 7 feet for habitable spaces and hallways, Section 1207.2). Additional checks addressed load path continuity (ensuring uninterrupted structural load transfer, Section 1604.4) and concrete slab thickness (minimum 6 inches with a vapor retarder, Section 1907.1). This assignment offered students a practical opportunity to integrate regulatory understanding with GenAI capabilities, bridging theoretical knowledge with automation and model interrogation in a real-world BIM environment.

*Survey Design*

Following the AI-driven assignment, students participated in a structured survey designed to thoroughly assess the educational impact of integrating GenAI into BIM education. This research study received approval from the University of Houston Institutional Review Boards (IRB) (STUDY00005289). The survey aimed to capture students' experiences comprehensively, focusing both on their subjective workload and their perceptions regarding the effectiveness and practicality of using AI tools. The survey was divided into three main parts:

The first section collected basic demographic and experiential information, including age group, gender identity, and years of construction-related experience. It also explored prior familiarity with BIM and code compliance procedures, as well as exposure to LLMs. These questions helped contextualize the responses by capturing the diversity of student backgrounds and baseline technological readiness.

To evaluate students' perceived workload systematically, the survey incorporated the NASA Task Load Index (NASA-TLX), a validated instrument for quantifying subjective workload during task performance (Hart & Staveland, 1988). This index measures workload across six dimensions: 1) Mental Demand: the cognitive effort required, including decision-making, memory, and problem-solving; 2) physical demand: related to the physical effort required to operate tools such as the keyboard and mouse; 3) temporal demand: the perceived time pressure perceived during the task; 4) Performance: Self-assessment of success and satisfaction in achieving task goals; 5) Effort: The mental and physical exertion required to maintain performance; and 6) Frustration level measures the extent of stress, insecurity, or annoyance experienced during the assignment. Each dimension was rated on a scale from low (1) to high (10), allowing for clear and standardized comparisons of workload across the various components. This quantitative approach enabled detailed insights into the specific workload areas influenced by introducing GenAI tools into the BIM rule-checking process.

The Third section of the survey contained questions specifically formulated to gauge student perceptions of the AI-driven compliance checking method. These questions explored several aspects critical to understanding the practical application and student acceptance of GenAI tools, including: Time Efficiency: Students evaluated how significantly they believed the AI method could reduce the time required for rule-checking tasks compared to traditional methods. Ease of Use and Effectiveness: Questions assessed how easily students could use the AI-generated Python scripts to identify non-compliant BIM elements, providing insights into usability and practical effectiveness. Openness to Technology Adoption: Students indicated their general openness to integrating new technologies such as GenAI into their academic and future professional practices. Overall Effectiveness Rating: Students provided an overall effectiveness rating of the AI-driven method, reflecting their comprehensive evaluation considering ease of use, robustness, and practical advantages compared to traditional manual approaches. Table 2 shows the sample questions from the survey.

*Student Interview*

Structured interviews were conducted with selected students following the AI-driven rule-checking assignment to gain deeper qualitative insights into their experiences. The primary objective of these interviews was to explore in greater detail the challenges faced by students, their interaction with GenAI tools, and their perceptions of the learning process. Key interview components included questions regarding the ease of use of GenAI, experiences with prompt engineering, and students' suggestions for future improvements in instruction and assignment design.

**Results**

*NASA-TLX Workload and Score*

The results of the NASA-TLX survey, along with the performance scores, offer a comprehensive understanding of students' cognitive, physical, and emotional engagement with the AI-powered code compliance checking assignment. As shown in Table 3, the highest reported workload dimension was Mental Demand (Mean = 7.36, SD = 1.98, Mode = 7), underscoring the substantial cognitive effort involved. This aligns with the assignment's requirements, where students engaged in interpreting building codes, formulating precise prompts, and refining AI-generated scripts, all demanding tasks that taxed memory, decision-making, and attention. A similarly high score was observed for Effort (Mean = 7.00, SD = 2.24, Mode = 7), confirming that students had to invest considerable mental and physical resources to successfully complete the tasks.

Other workload components were also rated notably. Physical Demand had a mean of 6.55 (SD = 2.35, Mode = 7), which, despite the virtual nature of the activity, likely reflects the sustained interaction with BIM and AI interfaces tasks that involved frequent manipulation of digital tools such as the Python shell in Revit. Temporal Demand was similarly moderate to high (Mean = 6.40, SD = 2.49, Mode = 7), indicating that students perceived the pace of the task and time limitations as significant challenges. Despite these demands, the self-reported Performance score remained relatively moderate (Mean = 6.47, SD = 2.52, Mode = 6), suggesting that while students generally felt they completed the task successfully, experiences varied, and some struggled to meet expectations. Frustration, meanwhile, had the highest variability (SD = 2.67), with a mean of 6.25 and mode of 8, indicating that while some students felt secure and gratified during the task, others experienced noticeable stress or irritation.

The overall group score across all six workload dimensions was 6.7, reflecting a moderately high level of perceived workload. This suggests that students found the assignment cognitively demanding and procedurally complex, yet still manageable with sustained effort.

On the other hand, the Assignment Score, which serves as an objective measure of performance, had a mean of 79.64, a standard deviation of 20.18, and a mode of 80. This average falls at the upper end of the B range, which generally reflects satisfactory to good performance. While most students demonstrated a reasonable grasp of the material and were able to complete the AI-driven compliance checks, the average does not reflect excellent performance. The variation in scores suggests that while some students performed well, others struggled, likely due to gaps in experience with Revit, Python scripting, or formulating effective AI prompts.

*Perceptions of Technology Adoption and Task Usability*

In addition to workload and performance measures, the study examined students' perceptions of the AI-powered code compliance checking method by focusing on three key categorical variables: openness to adopting new technology, the perceived time-saving benefits of the AI method, and the ease of identifying non-compliant elements using AI.

As shown in Table 4, a substantial majority of students expressed a willingness to embrace new technological tools, with 41.82% indicating they were open "To a moderate extent" and 38.18% "To a great extent." This finding highlights an overall positive attitude toward technological innovation in the learning process. However, a small group (10.91% "Very little" and 7.27% "To a small extent") remained hesitant, suggesting that not all students were fully comfortable adapting to emerging AI-driven tools, possibly due to unfamiliarity with coding environments or limited prior exposure to GenAI systems.

Perceptions of the AI method's impact on time efficiency were also largely positive. A clear majority of students (52.73%) believed that the method saved "A meaningful amount of time," and an additional 23.64% reported "A somewhat meaningful amount of time." In contrast, 9.09% indicated that it saved "Very little" time, with the same percentage (9.09%) selecting "Not at all." A smaller group (5.45%) perceived the time-saving benefit to be "An incomparable amount of time." These results suggest that although the AI-powered method offered significant time-saving potential for most students, some experienced limited benefits, likely reflecting individual differences in prompt formulation, interpretation of AI outputs, or navigation of the Revit-based workflow.

When asked about the ease of identifying non-compliant elements using the AI method, 36.36% of students selected "To a moderate extent," indicating a fair level of confidence in the tool's usability. A nearly equal proportion (34.55%) selected "To a small extent," while 16.36% reported "Very little" and 9.09% chose "Not at all." Only 3.64% of students rated the method as highly easy to use, selecting "To a great extent." These results highlight that while many students found the tool somewhat usable, a significant portion encountered difficulties. This underscores the need for clearer guidance, prompt templates, and instructional scaffolding to support effective use of AI in model checking tasks.

### *Regression Analysis*

To further explore the factors influencing students' experiences with AI-powered code compliance checking, multiple linear regression analyses were conducted using three key dependent variables: Frustration, Overall Effectiveness, and Openness to Use New

Technology. The goal was to examine how task demands and students' perceptions of the AI method influenced these outcomes. The results of the regression models are presented in Table 5.

Frustration Model: The model predicting frustration showed that effort, temporal demand, and mental demand were all significant positive predictors. Higher reported effort (beta = +0.860, p <.001), temporal demand (beta = +0.555, p <0.001), and mental demand (beta = +0.447, p =0.001) were associated with increased frustration. These results indicate that when students perceived the task as cognitively and physically demanding, they were more likely to experience negative emotions such as stress or irritation. At the same time, students who found the AI method helpful in identifying non-compliant elements (beta = –0.323, p =0.016) or believed it reduced the time spent on the task (beta = –0.299, p =0.027) reported lower levels of frustration. These findings suggest that students who viewed the AI tool as practical and time-efficient experienced fewer emotional difficulties while completing the task.

Overall Effectiveness Model: In the overall effectiveness model, frustration (beta = –0.596, p =0.001) and temporal demand (beta = –0.509, p =0.016) were negatively associated with students' ratings of how effective they found the AI method. This indicates that students who experienced less pressure and discomfort were more likely to view the method as effective. On the other hand, positive perceptions of the AI method were strong predictors of effectiveness. Students who felt that the AI tool saved time (beta = +0.521, p <.001) and made it easier to identify non-compliant elements (beta= +0.672, p <.001) rated it more favorably. These findings highlight that students' evaluations were shaped by practical outcomes, such as task speed and clarity.

Openness to Use New Technology Model: Frustration was a negative predictor (beta = –0.491, p =0.011), meaning students who felt more discouraged or stressed during the task were less open to using similar tools in the future. In contrast, students who believed the AI method reduced time (beta = +0.444, p =0.001), helped identify non-compliant elements (beta = +0.472, p <.001), and required meaningful mental engagement (beta = +0.376, p =0.027) were more open to using new technology. This suggests that both practical benefits and intellectual engagement contributed to students' willingness to adopt AI tools in the future.

***Student Reflections on the AI Experience***

In addition to quantitative survey responses, structured interviews and open-ended survey comments provided valuable insights into students' experiences with the AI-powered rule-checking task. A recurring theme was the iterative nature of working with GenAI. Several students noted that accurate code generation often required multiple prompt refinements. While they appreciated the automation AI offered, they emphasized the importance of detailed and precise prompts. As one student put it, "The more detailed your prompt is, the better results I was getting."

A second common point was the challenge of troubleshooting AI-generated scripts, especially for students without prior coding experience. Many expressed difficulties interpreting and modifying code when errors appeared, often relying on ChatGPT not just to generate scripts, but to debug and correct them. Students from non-computer science backgrounds emphasized their dependency on AI and their limited confidence in resolving issues independently.

Despite these challenges, most students acknowledged that AI significantly reduced the time needed to complete rule-checking tasks compared to manual approaches. They especially valued AI's speed in identifying non-compliant elements once the code worked. However, several students expressed skepticism toward the accuracy and reliability of AI-generated scripts, citing inconsistent results when using similar prompts or encountering misinterpretations in complex rules.

Finally, across both interviews and surveys, students repeatedly suggested that additional training, practical coding guidance, and example-driven instruction would have enhanced their learning experience. Specifically, they recommended tutorials on common AI errors, an explanation of Python basics relevant to the assignment, and improved scaffolding for first-time AI users.

**Interpretation of Findings**

The integrated results from survey analysis, regression models, and student reflections offer a multidimensional view of the learning experience associated with implementing an AI-powered rule-checking assignment in the BIM-related courses. These findings not only reflect students' perceptions but also point to specific instructional and design decisions that shaped those experiences.

One of the clearest outcomes was the high mental and temporal demand scores, paired with substantial reported effort. This mirrors the cognitive load and critical thinking challenges documented in a systematic review of ChatGPT in education (Lo et al., 2024). The load in our course is directly tied to its workflow: students had to craft prompts, interpret AI outputs, and iteratively debug scripts, an effort-intensive cycle. Recent LLM-based compliance checking study notes similar technical complexity and repeated refinement when rules are ambiguous or data are scarce. Interview quotes from our students ("I needed four or five prompt revisions before the script worked") echo that finding. This iterative prompting process, while pedagogically valuable for building understanding, likely contributed to increased workload and frustration, especially among students with limited coding backgrounds. Finally, opaque, trial-and-error AI workflows can heighten cognitive strain and user frustration unless accompanied by clear guidance (Yu & Guo, 2023), a pattern our NASA-TLX scores and qualitative feedback confirm, especially for participants with limited coding backgrounds

The regression results showed that effort, temporal demand, and mental demand were the strongest predictors of frustration, and interview comments confirmed that students who spent the most time debugging or rewriting prompts felt the greatest irritation. By contrast, those who believed the AI genuinely saved time and clearly flagged non-compliant elements reported noticeably lower frustration. Although educational studies have likewise tied heavy cognitive load to higher stress when AI tasks are poorly scaffolded (Lo et al., 2024; Yu & Guo, 2023), and (Lee et al., 2024). note that limited instructor guidance can compound this effect. However, our findings suggest that frustration was not driven by workload alone, but also by whether the AI tool was perceived as helpful in managing that workload.

Perceptions of the AI method's usefulness also played a strong role in how effective students found the method overall. Regression results showed that students who believed the tool saved time (beta =+0.521, p<.001) and made it easier to identify non-compliant elements (beta =+0.672, p<.001) rated it more favorably. This pattern mirrors prior work showing that learners adopt and value AI when it demonstrably increases efficiency and clarity (Chan & Hu, 2023; Ghimire et al., 2023; Yusuf et al., 2024). Conversely, higher temporal demand and frustration reduced perceived effectiveness an effect also reported in studies that link elevated workload or opaque AI behavior to diminished user trust and satisfaction (Lee et al., 2024; Yu & Guo, 2023). Together, our findings reinforce the broader evidence that practical, time saving benefits boost perceived effectiveness, whereas lingering cognitive strain undermines it.

A similar pattern emerged regarding students' openness to adopting new technology. Regression coefficients showed that students who viewed the AI workflow as time efficient and practically beneficial were markedly more willing to use similar tools in the future, an effect widely reported in studies where perceived usefulness and ease of application drive technology acceptance (Chan & Hu, 2023; Yusuf et al., 2024). Openness was also positively associated with higher mental engagement and with greater construction-related experience, echoing BIM education research that finds prior domain familiarity boosts receptiveness to novel digital methods (Meterelliyöz & Özener, 2022). However, while previous studies have highlighted the role of domain familiarity in fostering openness to innovation, the observed link between openness and elevated mental engagement appears to be a novel contribution not explicitly addressed in earlier research. In contrast, frustration carried a significant negative weight, indicating that emotional discomfort can dampen future adoption unless well targeted support is provided, which is consistent with findings that high cognitive strain or opaque AI behavior undermines user confidence (Lo et al., 2024; Yu & Guo, 2023).

   Despite the reported mental effort and moderate frustration, most students completed the assignment. However, the average assignment score of approximately 80% falls short of the level typically associated with strong or excellent performance in academic settings. While this result indicates that students were generally able to engage with the assignment and reach acceptable outcomes, it also suggests that the learning goals were not fully achieved across the board. The assignment was designed not only to introduce AI tools but also to develop deeper skills in prompt formulation, error handling, and rule interpretation. Given that 80% represents the lower end of the 'B' range, there is clear room for improvement. This outcome may reflect the challenges students faced in adapting to unfamiliar tools, such as Python scripting in Revit, and the cognitive load involved in debugging AI-generated outputs. These persistent technical hurdles emphasize the continued necessity for structured instructional support, foundational coding education, and iterative tool refinement.

Finally, the qualitative feedback offered several suggestions that align with the statistical findings. Students recommended the inclusion of structured examples, introducing more pre-written prompt templates, and a walkthrough of Python syntax. These suggestions reflect a broader recognition that while GenAI tools can enhance learning and streamline technical tasks, their success depends heavily on how well they are introduced, supported, and integrated into the course structure.

**Lessons Learned from the Implementation of AI-Driven Rule-Checking**

Based on the survey results, regression analysis, and student reflections, several key lessons emerged from the implementation of the AI-driven tool in a BIM rule-checking educational context. These lessons offer practical guidance for future iterations of integrating GenAI into construction management education.

Refine technology before introducing it to students: Although the AI-driven system showed promising results in assisting students with code compliance checking, the findings highlight the importance of ensuring that such tools are sufficiently mature and user-friendly before being deployed in a classroom setting. Students encountered inconsistencies in output, required multiple prompt refinements, and often relied on trial-and-error debugging. These issues increased cognitive load and contributed to frustration, as reflected in both the NASA-TLX ratings and the regression findings. When students perceive tools as unreliable or opaque, their confidence and engagement can decline. A key takeaway is that educational AI tools must be carefully tested, streamlined, and improved for usability and reliability prior to integration into student assignments.

Allocate more time for practice and skill applications: While the instructional session provided detailed content on prompt engineering and the structure of AI-driven, student feedback clearly indicated that the time allocated for hands-on practice was insufficient. The learning curve associated with AI prompting, particularly when tied to technical outputs like Python scripts, requires iterative engagement. Future implementations should dedicate more in-class or workshop time for guided experimentation, where students can receive feedback as they refine prompts and navigate AI outputs. Expanding time for practice could also improve students' openness to new technologies by reducing frustration and building confidence through supported trial-and-error.

Introduce basic coding concepts early in the process: The assignment required students to interact with Python scripts, yet no formal instruction was provided on what coding is, how Python works, or even basic error interpretation. While the course was not designed to teach programming, the lack of foundational coding knowledge left many students overly dependent on AI without understanding the logic behind the scripts. This gap may have contributed to the modest average assignment score, as students were often completing tasks without fully understanding the reasoning behind them. A key lesson here is that students benefit greatly from even minimal exposure to fundamental programming concepts, such as syntax, variables, and debugging principles, especially when AI-generated outputs rely on them.

Emphasize productive use of AI as a learning partner: Rather than using AI merely as a script generator, students should be encouraged to use it interactively, asking it to explain errors, clarify logic, or justify code structure. The current model allowed students to repeatedly copy, paste, and adjust prompts without necessarily building deeper understanding. Embedding instructional guidance on how to use GenAI tools not just to generate answers, but to learn from the process, can significantly enhance their educational value.

AI should be framed as an assistant, not a replacement: A recurring insight from both the quantitative and qualitative data is that students began to view the AI not as a complete solution, but as a technical assistant. They recognized its value in speeding up repetitive tasks but also understood that the outputs required careful human interpretation and oversight. This is a critical conceptual shift. Students must be trained to view GenAI tools in technical domains not as decision-makers, but as collaborative partners whose outputs require refinement and validation. This mindset fosters accountability and aligns with the evolving role of AI in professional construction settings. This aligns with both the survey data and student reflections, where many emphasized the need to review, correct, and validate AI-generated outputs before trusting them for real-world applications.

**Limitations and Future Directions**

While this study highlights the promising role of GenAI in BIM education, several limitations must be acknowledged. First, the study was conducted across two semesters with a combined sample of 55 students, all enrolled in the same courses at a single institution. While this provides more robust data than a single-semester sample, it still limits the generalizability of the findings. The results may not fully reflect diverse instructional contexts, student backgrounds, or institutional settings. Long-term impacts on learning outcomes, such as skill retention and transferability across projects, were also not assessed. Second, the absence of a control group limits the ability to isolate the specific effects of AI assistance. All students in the study used the AI-driven method, and no comparison was made to a manual-only rule-checking group. A direct comparison would strengthen future evaluations by offering a clearer picture of the added value of AI tools in construction education. Third, although the AI-driven method was designed to be accessible without programming expertise, students encountered several usability challenges. These included inconsistent AI outputs, vague error messages, and hallucinations, particularly when handling ambiguous or complex building codes. These issues contributed to cognitive load and frustration, especially for students without prior experience using Python. While prompt engineering lowered the barrier to entry, effective use still required an ability to iterate, evaluate, and troubleshoot the AI-generated scripts. Finally, the AI-driven tool itself remains under development. Although it demonstrated strong potential to assist in rule-checking tasks, its reliability, usability, and consistency need further improvement. Students' experiences suggest that continued refinement and testing are critical before broader adoption.

Based on these limitations, the following recommendations are proposed for future research: 1) Include a control group and expand the sample. Future studies should include a manual rule-checking group for comparison and recruit a more diverse student population across courses or institutions to improve generalization. 2) Conduct longitudinal studies and Evaluate the long-term impact of AI-assisted learning on skills retention, independent AI use, and transfer of knowledge across projects or courses to better understand sustainable integration into construction education.

Building upon the lessons learned, future research should also: 1) Introduce foundational instruction earlier. As previously noted, early curriculum integration of GenAI concepts, prompt engineering techniques, and basic Python programming principles could mitigate learning challenges and enhance student engagement. 2) Provide structured support resources. As recommended in the lessons learned, additional prompt templates, error-handling guides, and step-by-step walkthroughs for

interpreting AI-generated code could significantly support learners with limited technical backgrounds

. 3) Further refine the AI-driven tool based on student feedback, emphasizing improvements in reliability, consistency of outputs, and interface usability to facilitate broader classroom adoption.

.

**Conclusion**

This study examined the integration of GenAI into a BIM course through an AI-driven code compliance checking assignment, with the goal of extracting lessons learned for future implementation. The findings revealed that while AI tools like AI-driven offer strong potential to enhance efficiency and support learning, their success in educational settings depends heavily on usability, instructional support, and alignment with students' prior experience.

Students demonstrated a willingness to adopt new technologies and recognized the time-saving benefits of AI, but also faced cognitive challenges, frustration, and inconsistent results, particularly when lacking prior exposure to coding or GenAI. These experiences highlighted the importance of improving the AI tool before classroom deployment, introducing basic technical concepts early, and emphasizing the proper use of AI as a learning partner rather than a shortcut.

The combination of survey analysis, statistical testing, and student interviews provided a well-rounded understanding of how GenAI tools can both support and challenge learners. Key recommendations include refining the tool's usability, expanding AI-related training, and incorporating structured support materials to reduce cognitive load and increase confidence.

Ultimately, this study affirms that GenAI has meaningful potential in construction education but only when paired with thoughtful instructional design, proper scaffolding, and continuous refinement based on student feedback. These insights offer a practical foundation for improving AI integration in BIM-based courses and guiding future educational innovations in the field.

**Declaration of generative AI and AI-assisted technologies in the writing process**

During the preparation of this work, the authors used ChatGPT in order to proofread the manuscript. After using this tool, the authors reviewed and edited the content as needed and take full responsibility for the content of the publication.

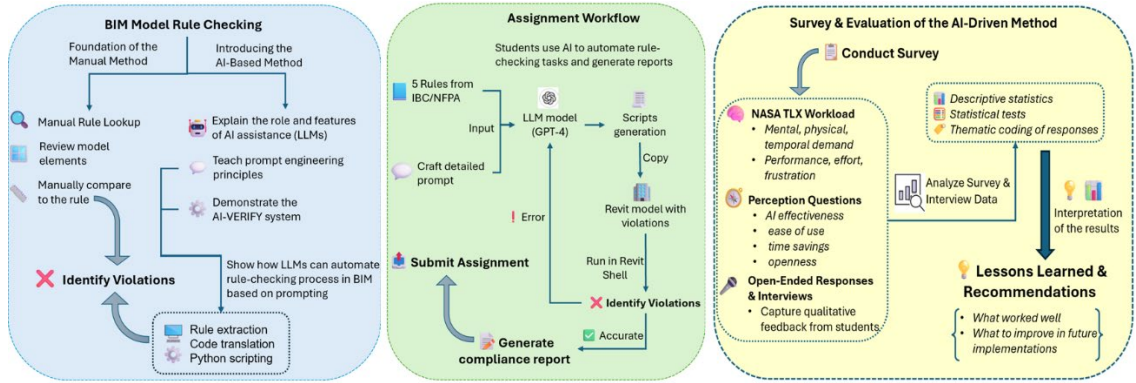

figure 1: Overview of the AI-Driven Code Compliance Lecture Workflow and Evaluation Methodology

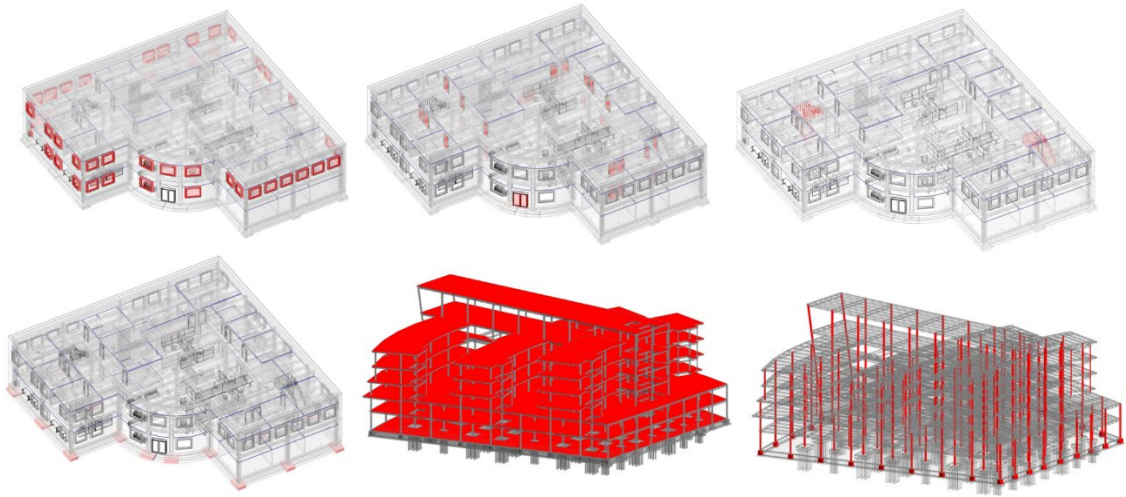

Figure 2: BIM Elements Highlighted for Code Compliance Verification Tasks Based on IBC 2018 and NFPA 101-2024 (Pre-Checking View).

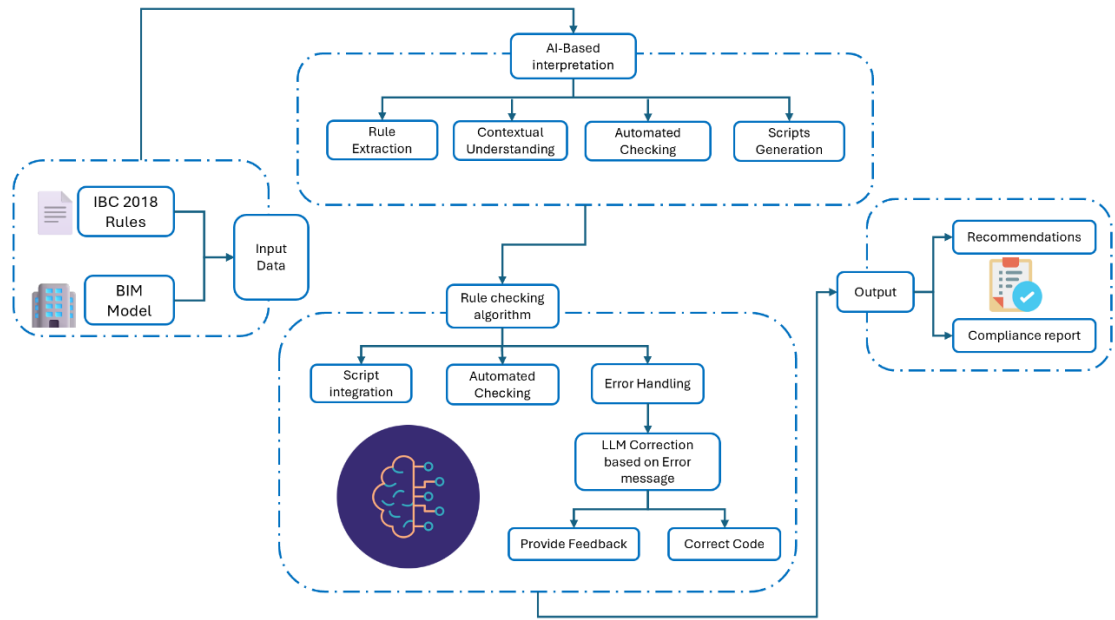

Figure 3: AI-Based Framework for Rule Checking in BIM Model

Table 1: Structured Prompt and Follow-Up Strategy for Generating Python Code to Check Stair Compliance in Revit

| Step | Prompt Segment | Purpose |
|---|---|---|
| | First prompt | |
| 1. Role Assignment | "You are a building code compliance expert with BIM and Revit Python Shell experience." | Establishes the AI's role and area of expertise to get context-aware responses. |
| 2. Task Definition | "Interpret the rule above and generate a Python script that checks each stair element in a Revit 2024 model." | Clearly states what the AI is expected to do: interpret and generate code. |
| 3. Method Transparency | "Clearly describe how the script evaluates riser height and tread depth, for example, whether it checks values from family parameters, geometry, or element properties." | Instructs the AI to explain the approach it uses, which improves traceability and review. |
| 4. Output Constraint | "The output should include only stairs that do not meet the requirements, along with their measured values." | Limits the result to non-compliant cases for efficiency and clarity. |
| Complete prompt | "You are a building code compliance expert with BIM and Revit Python Shell experience. Interpret the rule above and generate a Python script that checks each stair in a Revit 2024 model. Clearly describe how the script evaluates riser height and tread depth, for example, whether it reads from family parameters, object geometry, or properties. The output should include only stairs that do not meet the requirements, along with their measured values. " | |
| Follow-Up Prompt | | |
| 1. Error Input (User Action) | [User pastes the error message or describes the incorrect output] | This provides the AI with context about what went wrong in the previous result. |
| 2. Diagnostic Request | "Please analyze this issue and explain the possible causes." | Asks the AI to assess the failure and identify why the original script didn't work correctly. |
| 3. Solution Request | "Suggest solutions and tell me what might be wrong in the current approach." | Prompts the AI to provide corrections or adjustments to improve the output. |

| 4. Clarification Invitation | "Also, if you need more details from the model (e.g., parameter names or values), ask for them so you can revise the script accurately." | Gives the AI permission to ask for missing information it needs before suggesting a fix. |

Table 2. Sample Questions from the Survey

| Question | Scale/Options |
| --- | --- |
| How much mental effort was required in each method? | 1 (Low) to 10 (High) |
| How effective was the AI-powered method compared to manual processes? | 1 (Not effective at all) to 10 (Highly effective) |
| How open are you to adopting new technology for future tasks? | Not at all, very little, To a small extent, To a moderate extent, To a great extent |

Table 3: Descriptive Statistics of NASA-TLX Dimensions and Assignment Score

| Category | Mean | STD | Mode |
| --- | --- | --- | --- |
| Mental Demand | 7.36 | 7.0 | 7.0 |
| Physical Demand | 6.55 | 7.0 | 7.0 |
| Temporal Demand | 6.40 | 7.0 | 7.0 |
| Performance | 6.47 | 7.0 | 6.0 |
| Effort | 7.00 | 7.0 | 7.0 |
| Frustration | 6.25 | 6.0 | 8.0 |
| Assignment Score | 79.64 | 80.0 | 80.0 |

Table 4: Student Responses on Technology Adoption, Time-saving, and Ease of Use

| Variable | Response Category | Frequency | Percentage |
| --- | --- | --- | --- |
| Openness to Use New Technology | To a moderate extent | 23 | 41.82% |
| | To a great extent | 21 | 38.18% |
| | Very little | 6 | 10.91% |
| | To a small extent | 4 | 7.27% |
| | Not at all | 1 | 1.82% |

| Variable | Response Category | Frequency | Percentage |
|---|---|---|---|
| AI Method Saves Time | A meaningful amount of time | 29 | 52.73% |
| | A somewhat meaningful amount of time | 13 | 23.64% |
| | Very little | 5 | 9.09% |
| | Not at all | 5 | 9.09% |
| | An incomparable amount of time | 3 | 5.45% |
| Ease of Identifying Non-Compliant Elements | To a moderate extent | 20 | 36.36% |
| | To a small extent | 19 | 34.55% |
| | Very little | 9 | 16.36% |
| | Not at all | 5 | 9.09% |
| | To a great extent | 2 | 3.64% |

Table 5: Regression Results for Frustration, Overall Effectiveness, and Openness to Use New Technology

| Dependent | Predictor | B | SE | Std. Beta | t | p |
|---|---|---|---|---|---|---|
| Frustration | | | | | | |
| | Effort | 1.03 | 0.18 | 0.86 | 5.69 | 0.000 |
| | Temporal demand | 0.59 | 0.12 | 0.55 | 4.86 | 0.000 |
| | Mental demand | 0.6 | 0.16 | 0.44 | 3.64 | 0.001 |
| | Easily identify non-compliant elements using AI methods | -0.93 | 0.33 | –0.323 | –2.48 | 0.016 |
| | AI-powered methods can reduce the time spent | -0.81 | 0.35 | –0.29 | –2.27 | 0.027 |
| Overall effectiveness | | | | | | |
| | Frustration | –0.50 | 0.14 | –0.59 | –3.42 | 0.001 |
| | Temporal demand | –0.45 | 0.18 | –0.50 | –2.51 | 0.016 |
| | AI-powered methods can reduce the time spent on the same process | 1.19 | 0.26 | 0.52 | 4.44 | 0.000 |
| | Easily identify non-compliant elements using AI methods | 1.51 | 0.22 | +0.67 | 6.61 | 0.000 |
| Open to using a new technology | | | | | | |
| | Years of experience | 0.44 | 0.13 | +0.37 | 3.33 | 0.002 |
| | Frustration | –0.18 | 0.06 | –0.49 | –2.62 | 0.011 |
| | Mental demand | 0.18 | 0.08 | +0.37 | 2.27 | 0.027 |
| | AI-powered methods can reduce the time spent | 0.44 | 0.12 | +0.44 | 3.60 | 0.001 |
| | Easily identify non-compliant elements using AI | 0.46 | 0.11 | +0.47 | 3.90 | 0.000 |